\documentclass[preprint,10pt,a4paper,twocolumn,tightenlines]{revtex4}%
\usepackage{amsfonts}
\usepackage{amsmath}
\usepackage{amssymb}
\usepackage{graphicx}%
\providecommand{\U}[1]{\protect\rule{.1in}{.1in}}

\begin{document}
\title{Decoherence-free evolution of time-dependent superposition states of two-level
systems and thermal effects}
\author{F. O. Prado$^{1}$, N. G. de Almeida$^{2}$, E. I. Duzzioni$^{1}$, M. H. Y.
Moussa$^{3}$, and C. J. Villas-Boas$^{4}$}
\affiliation{$^{1}$Universidade Federal de Uberl\^{a}ndia, Caixa Postal 593, 38400-902,
Uberl\^{a}ndia,\textit{ MG}, Brazil}
\affiliation{$^{2}$Instituto de F\'{\i}sica, Universidade Federal de Goi\'{a}s, 74001-970,
Goi\^{a}nia, \textit{GO}, Brazil}
\affiliation{$^{3}$Instituto de F\'{\i}sica de S\~{a}o Carlos, Universidade de S\~{a}o
Paulo, Caixa Postal 369, 13560-970, S\~{a}o Carlos, \textit{SP}, Brazil}
\affiliation{$^{4}$Departamento de F\'{\i}sica, Universidade Federal de S\~{a}o Carlos,
P.O. Box 676, 13565-905, S\~{a}o Carlos\textit{, SP, }Brazil}

\begin{abstract}
In this paper we detail some results advanced in a recent letter [Phys. Rev.
Lett. \textbf{102}, 073008 (2009) ] showing how to engineer reservoirs for
two-level systems at absolute zero by means of a time-dependent master
equation leading to a nonstationary superposition equilibrium state. We also
present a general recipe showing how to build nonadiabatic coherent evolutions
of a fermionic system interacting with a bosonic mode and investigate the
influence of thermal reservoirs at finite temperature on the fidelity of the
protected superposition state. Our analytical results are supported by
numerical analysis of the full Hamiltonian model.

\end{abstract}

\maketitle

\section{Introduction}

The purpose of the engineering reservoir program \cite{Poyatos} is to protect
a specific state against the decoherence stemming from the natural coupling
between a quantum system and the reservoir. To engineer a reservoir, a given
system, whose state is to be protected, is compelled to engage in additional
interactions besides that with the natural reservoir. The engineering
reservoir technique is then applied to make these additional interactions
prevail, modifying the dissipative Liouvillian in a specific way that drives
the system to the desired equilibrium with the engineered reservoir. Rightly
connected to the engineering Hamiltonian program \cite{Fabiano}, the reservoir
engineering has been developed for trapped ions \cite{Ions, Matos} and atomic
two-level ($TL$) systems \cite{Atoms}. Recently, under the assumption of a
squeezed engineered reservoir, a way to observe the adiabatic geometric phase
acquired by a protected state has been proposed \cite{Carollo2006a,
Carollo2006c, Yin 2007} under the assumption of an adiabatic evolution of the
system-reservoir parameters. It is worth mentioning recent results on the
possibility to drive an open many-body quantum system into a given pure state
by an appropriate design of the system-reservoir coupling \cite{Diehl1,Diehl2}.

In this paper we detail the general recipe presented in
Ref.\cite{PRLFabiano09} showing how to build nonadiabatic coherent evolutions
of a two-level ion trapped into a leaky cavity, in contrast with the
applications of the engineering reservoir technique which require adiabatic
conditions \cite{Carollo2006a, Carollo2006c, Yin 2007}. Furthermore, we show
how to implement the proposal of Ref. \cite{Carollo2006b} for controlling the
slow changes of a protected $TL$ system through the parameters of the
engineered reservoir. We also investigate the influence of thermal reservoirs
at finite temperature on the fidelity of the protected superposition state.

This paper is organized as follows. In Section II we briefly present the
engineering reservoir program, including the advances presented in Ref.
\cite{PRLFabiano09}. In Section III we show in details a general recipe to
engineer reservoirs for\ a fermionic system interacting with a bosonic mode.
As an application we derive an effective reservoir for a time-dependent
superposition state of the internal degrees of freedom of an ion trapped into
a leaky cavity. In Section IV we present numerical results supporting our
approximations and an analysis of the temperature effects on the protected
time-dependent superposition. Finally, in Section V we present our conclusions.

\section{The extended engineering reservoir program}

Before we present the technique to protect superpositions of quantum states
evolving nonadiabatically, we review basics results concerning the engineering
reservoir program. To this end, we focus our attention on a system interacting
with a reservoir, both system and reservoir being modeled by harmonic
oscillators. The coupling, as usual, is assumed to be linear in
position-position. From this model one can deduce a master equation, obtained
by tracing over the reservoir variables, to study the evolution of the single
harmonic oscillator. By driving the system with external fields and/or
allowing the system to interact with additional systems, each one possessing
its own reservoir, it is possible to obtain, through some approximations, an
effective master equation which is the starting point to protect a given state
by means of engineered reservoirs, as we detail in the following.

\subsection{Protection of a stationary quantum state}

The master equation describing the evolution of the density operator $\rho$ of
a given system coupled with its natural reservoir in the interaction picture
and at zero temperature is%
\begin{equation}
\overset{\cdot}{\rho}=\frac{\Gamma}{2}\left(  2a\rho a^{\dagger}-a^{\dagger
}a\rho-\rho a^{\dagger}a\right)  , \label{eqmestra0}%
\end{equation}
where $a^{\dagger}$ ($a$) is the creation (annihilation) operator in the Fock
states of the system. As it is well known, the only steady state resulting
from Eq.(\ref{eqmestra0}) is the vacuum $\left\vert 0\right\rangle $, which is
an eigenstate of the annihilation operator with zero eigenvalue: $a\left\vert
0\right\rangle =\alpha\left\vert 0\right\rangle $, $\alpha=0$. The main goal
of the standard engineering reservoir \cite{Poyatos} is to obtain, in the
interaction picture and at zero temperature, an effective master equation in
the form
\begin{equation}
\overset{\cdot}{\rho}=\frac{\Gamma}{2}\left(  2\mathcal{O}\rho\mathcal{O}%
^{\dagger}-\mathcal{O}^{\dagger}\mathcal{O}\rho-\rho\mathcal{O}^{\dagger
}\mathcal{O}\right)  , \label{eqmestra1}%
\end{equation}
where $\Gamma$ is the effective decay rate of the engineered reservoir which
is coupled to the quantum system in a specific way characterized by the
time-independent system operator $O$. Proceeding in analogy with
Eq.(\ref{eqmestra0}), the only pure steady state of this system is the
eigenstate $\left\vert \psi\right\rangle $ of the operator $O$ with null
eigenvalue, ensuring that there is no further eigenstate $\left\vert
\phi\right\rangle $ of $O$ such that $\left[  \mathcal{O},\mathcal{O}%
^{\dagger}\right]  \left\vert \phi\right\rangle =0$. As a consequence,
$\left\vert \psi\right\rangle $ is the asymptotic state of the system
\cite{Matos}.

\subsection{Protection of a time-dependent quantum state}

The Ref. \cite{PRLFabiano09} adds an improvement on the engineering reservoir
program by allowing to remove the adiabatic constraint in the decoherence-free
evolution mentioned above. To understand this simple, yet effective
improvement, here we review the theory developed in Ref. \cite{PRLFabiano09}.
Consider the engineered time-dependent master equation in the interaction
picture
\begin{align}
\overset{\cdot}{\rho}  &  =-i\left[  H\left(  t\right)  ,\rho\right]
+\frac{\Gamma}{2}\left[  2\mathcal{O}\left(  t\right)  \rho\mathcal{O}%
^{\dagger}\left(  t\right)  -\mathcal{O}^{\dagger}\left(  t\right)
\mathcal{O}\left(  t\right)  \rho\right. \nonumber\\
&  \left.  -\rho\mathcal{O}^{\dagger}\left(  t\right)  \mathcal{O}\left(
t\right)  \right]  , \label{eqmestra2}%
\end{align}
where the Hermitian Hamiltonian $H\left(  t\right)  $ must be engineered
aiming to synthesize ---without adiabatic impositions--- the desired
time-dependence of the operator $\mathcal{O}\left(  t\right)  =R\left(
t\right)  \mathcal{O}_{R}R^{\dag}\left(  t\right)  $, with $\overset{\cdot
}{\mathcal{O}}_{R}=0$, $R\left(  t\right)  =T\exp\left(  -i\int_{0}%
^{t}H\left(  t^{\prime}\right)  dt^{\prime}\right)  $, and $T$ being the
time-ordering operator. Such a relation between $\mathcal{O}\left(  t\right)
$ and $R(t)$ justifies the above mentioned intimate connection between both
programs of engineering Hamiltonians and reservoirs; in fact, the reservoir
engineering technique relies on engineered Hamiltonians.\textbf{ }We stress
that we have achieved the time evolution of Eq. (\ref{eqmestra2}) in a
particular way that the engineered\ Hamiltonian $H\left(  t\right)  $ prompts
the specific operator $O\left(  t\right)  $ and its protected evolving
eigenstate $\left\vert \psi\left(  t\right)  \right\rangle $ (with null
eigenvalue). The key feature to be noted here is that, through the unitary
transformation $R\left(  t\right)  $, we recover the time-independent form of
the master equation:%
\begin{equation}
\overset{\cdot}{\rho_{R}}=\frac{\Gamma}{2}\left(  2\mathcal{O}_{R}\rho
_{R}\mathcal{O}_{R}^{\dagger}-\mathcal{O}_{R}^{\dagger}\mathcal{O}_{R}\rho
_{R}-\rho_{R}\mathcal{O}_{R}^{\dagger}\mathcal{O}_{R}\right)  .
\label{eqmestra3}%
\end{equation}
Therefore, the protected stationary state, $\left\vert \psi_{R}\right\rangle $
($\mathcal{O}_{R}\left\vert \psi_{R}\right\rangle =0$), turns out to be a
nonstationary state $\left\vert \psi(t)\right\rangle =R\left(  t\right)
\left\vert \psi_{R}\right\rangle $ in the original interaction picture of
(\ref{eqmestra2}), where $H\left(  t\right)  =i\overset{\cdot}{R}(t)R^{\dag
}(t)$. Getting rid of the adiabatic constraints, we are thus allowed to
manipulate the evolution of the protected state $\left\vert \psi
(t)\right\rangle $ through appropriate engineered Hamiltonian and reservoir.
It is worth mention that the time dependence of the protected state is closely
related to the properties of $H(t)$ and $O\left(  t\right)  $. If $\left[
H(t),\mathcal{O}\left(  t\right)  \right]  =\left[  i\overset{\cdot}%
{R}(t)R^{\dag}(t),\mathcal{O}\left(  t\right)  \right]  =i$ $\overset{\cdot
}{\mathcal{O}}\left(  t\right)  =0$, then it is straightforward to see that
$\left\vert \psi(t)\right\rangle =\left\vert \psi_{R}\right\rangle $, i.e.,
$\left\vert \psi(t)\right\rangle \,$is stationary apart from a global phase factor.

Next we develop a general theory of reservoir engineering for a system
interacting with a bosonic mode, i. e., we show how to use the bosonic decay
to build up the general master equation (\ref{eqmestra1})\textbf{. }Then we
apply this theory to protect quantum states of a $TL$ system, showing how to
obtain an effective interaction between this system and a cavity mode which
leads to the desired state protection mechanism.

\section{Reservoir engineering for a fermionic system interacting with a
dissipative bosonic field}

The starting point to develop a general theory of reservoir engineering for a
fermionic system (from now on "the system") interacting with a dissipative
bosonic field is the engineered effective Hamiltonian%
\begin{equation}
H_{eff}=\lambda_{eff}\left(  \mathfrak{O}a^{\dagger}+\mathfrak{O}^{\dagger
}a\right)  , \label{Hef1}%
\end{equation}
where $a^{\dagger}$ ($a$) is the creation (annihilation) operator of the
bosonic field, while $\mathfrak{O}$ and $\mathfrak{O}^{\dagger}$ are operators
associated with the system whose state we wish to protect, and $\lambda
_{eff}\ $is the effective coupling between the bosonic mode and the system. As
an example, some of us showed how to build up bimodal interactions in cavity
quantum electrodynamics using three- or two-level atoms \cite{Fabiano}. We
note that the interaction (\ref{Hef1}) is a bilinear form similar to the
interaction of a bosonic field with a natural reservoir which leads to the
well-know Liouvillian describing the amplitude damping mechanism. Thus, the
engineered interaction (\ref{Hef1}) must generate an effective dissipative
Liouvillian which competes with the natural one. When the engineered decay
rate is significantly larger than the natural one, the effective dissipative
Liouvillian will govern the dynamics of the system.

Once achieved the important step of building the interaction between the
system of interest and the quantized bosonic mode, we now add the dissipative
mechanism of the bosonic mode through the master equation
\begin{align}
\overset{\cdot}{\rho^{(sa)}}  &  =-i\left[  H_{eff},\rho^{(sa)}\right]
\nonumber\\
&  +\frac{\kappa}{2}\left(  2a\rho^{(sa)}a^{\dagger}-a^{\dagger}a\rho
^{(sa)}-\rho^{(sa)}a^{\dagger}a\right)  , \label{master-geral}%
\end{align}
written in the same representation we obtained $H_{eff}$; the factor $\kappa$
stands for the natural decay rate of the bosonic mode and the supra index in
$\rho^{(sa)}$ indicates its dependence from both the field ($a$) and system
($s$) operators. From the above equation we straightforwardly derive the
evolution equation for the matrix elements $\rho_{nm}^{(s)}=\left\langle
n\right\vert \rho^{(sa)}\left\vert m\right\rangle $ in the Fock basis:
\begin{align}
\overset{\cdot}{\rho}_{n,m}^{(s)}  &  =-i\lambda_{eff}\left[  \sqrt
{n}\mathfrak{O}\rho_{n-1,m}^{(s)}+\sqrt{n+1}\mathfrak{O}^{\dagger}\rho
_{n+1,m}^{(s)}\right. \nonumber\\
&  \left.  -\left(  \sqrt{m+1}\rho_{n,m+1}^{(s)}\mathfrak{O}+\sqrt{m}%
\rho_{n,m-1}^{(s)}\mathfrak{O}^{\dagger}\right)  \right] \label{master1}\\
&  +\kappa\sqrt{\left(  n+1\right)  \left(  m+1\right)  }\rho_{n+1,m+1}%
^{(s)}-\frac{\kappa}{2}\left(  n+m\right)  \rho_{n,m}^{(s)}.\nonumber
\end{align}
To engineer the reservoir, we assume that the decay constant $\kappa$ of the
bosonic mode is significantly larger than the effective coupling\textbf{
}$\lambda_{eff}$ in Eq. (\ref{master1}). This condition is easily achieved for
cavities with low quality factor $Q=\omega/\kappa$, where $\omega$ is the
bosonic mode frequency. Together with the good approximation of a reservoir at
zero temperature, this regime enables to consider only the matrix elements
$\rho_{mn}^{(s)}$ inside the subspace $\left\{  \left\vert 0\right\rangle
\text{,}\left\vert 1\right\rangle \right\}  $ of Fock states. Actually, Eq.
(\ref{master1}) can very well be described by the set of equations%

\begin{gather}
\overset{\cdot}{\rho}_{0,0}^{(s)}=-i\lambda_{ef}\left(  \mathfrak{O}^{\dagger
}\rho_{1,0}^{(s)}-\rho_{0,1}^{(s)}\mathfrak{O}\right)  +\kappa\rho_{1,1}%
^{(s)},\\
\overset{\cdot}{\rho^{(s)}}_{1,0}=\left(  \overset{\cdot}{\rho}_{0,1}\right)
^{\dagger}=-i\lambda_{ef}\left(  \mathfrak{O}\rho_{0,0}^{(s)}-\rho
_{1,1}\mathfrak{O}\right)  -\frac{\kappa}{2}\rho_{1,0}^{(s)},\\
\overset{\cdot}{\rho}_{1,1}^{(s)}=-i\lambda_{ef}\left(  \mathfrak{O}\rho
_{0,1}^{(s)}-\rho_{1,0}^{(s)}\mathfrak{O}^{\dagger}\right)  -\kappa\rho
_{1,1}^{(s)},
\end{gather}
which are similar to those written in the atomic basis in the reservoir
engineering program for trapped ions \cite{Matos}. The strong decay rate
$\kappa$ enables the adiabatic elimination of the elements $\rho_{01}$ and
$\rho_{11}$ from the equations above. The formal adiabatic elimination is
equivalent to assume $\overset{\cdot}{\rho^{(s)}}_{1,0}\simeq0$, allowing us
to write $\rho_{1,0}^{(s)}=-i2\frac{\lambda_{ef}}{\kappa}\left(
\mathfrak{O}\rho_{0,0}^{(s)}-\rho_{1,1}\mathfrak{O}\right)  $. We can thus
eliminate both $\rho_{1,0}^{(s)}$ and $\widetilde{\rho}_{0,1}^{(s)}$ from the
dynamics of the diagonal elements of the density matrix by substituting them
in the above equations. Since $\rho_{0,0}^{(s)}+\rho_{1,1}^{(s)}\simeq
\rho^{(s)}\equiv\rho_{sys}$, the reduced density operator for the system of
interest thus results%
\begin{equation}
\overset{\cdot}{\rho_{sys}}=\frac{\Gamma_{eng}}{2}\left(  2\mathfrak{O}%
\rho_{sys}\mathfrak{O}^{\dag}-\mathfrak{O}^{\dagger}\mathfrak{O}\rho
_{sys}-\rho_{sys}\mathfrak{O}^{\dagger}\mathfrak{O}\right)  \text{,}
\label{eff-master}%
\end{equation}
where $\Gamma_{eng}=4\lambda_{eff}^{2}/\kappa$ represents the effective decay
rate corresponding to the engineered reservoir. With this result we can easily
see that the reservoir engineering technique depends basically on the
manipulation of the interaction between the bosonic mode and the system whose
state we wish to protect.

The two kinds of master equations presented above, (\ref{eqmestra1}) and
(\ref{eqmestra2}), may be recovered through Eq. (\ref{eff-master}) by
identifying the operator $\mathfrak{O}$ with $O$ and $O_{R}$, respectively.
Such correspondence is established when the operator $O$ is written in the
usual interaction picture, while $O_{R}$ is written in an arbitrary
representation where it is stationary and Eq. (\ref{eqmestra2}) remains valid.
Below we show how to engineer a reservoir which allows to protect an arbitrary
evolving superposition state of the internal degrees of freedom of a trapped
ion/atom without the adiabatic constraints.

\subsection{Engineering reservoirs for a two-level system in a leaky cavity}

To implement the ideas discussed in the previous Section, we use a two-level
($TL$) system (s) characterized by the transition frequency $\omega_{0}$
between the ground $\left\vert g\right\rangle $ and excited $\left\vert
e\right\rangle $ states. This system can be either a $TL$ neutral atom in a
dipole trap \cite{rempe-kimble} or a $TL$ ion in a harmonic trap with
frequency $\nu$ \cite{blatt}. The transition $\left\vert g\right\rangle $
$\leftrightarrow$ $\left\vert e\right\rangle $ is driven by a classical field
of frequency $\omega_{c}$, with coupling strength $\Omega_{c}$. The $TL$
system is made to interact with a cavity mode field\textbf{ (}$a$\textbf{)} of
frequency $\omega_{a}$ under the Jaynes-Cummings Hamiltonian with Rabi
frequency $g$. Assuming the case of a two-level trapped ion, the Hamiltonian
modelling this system is given, within the rotating-wave approximation
($RWA$), by
\begin{gather}
H=\omega_{a}a^{\dagger}a+\omega_{0}\sigma_{z}/2+\nu b^{\dagger}b+\left\{
g\cos\left(  \overrightarrow{k}.\overrightarrow{x}\right)  a\sigma_{eg}\right.
\nonumber\\
\left.  +\Omega_{c}\operatorname*{e}\nolimits^{i\left(  \overrightarrow{k}%
_{c}.\overrightarrow{x}+\phi_{c}-\omega_{c}t\right)  }\sigma_{eg}%
+\mathrm{H{.c.}}\right\}  , \label{1}%
\end{gather}
where $b^{\dagger\text{ }}$($b$) is the creation (annihilation) operator of
the vibrational mode whose position operator is $\overrightarrow
{x}=(b^{\dagger}+b)/\sqrt{2m\nu}\widehat{x}$, $m$ being the ionic mass and
$\widehat{x}$ the unit vector along the vibrational direction. The wave
vectors $\overrightarrow{k}$ and $\overrightarrow{k}_{c}$ stand for the cavity
mode and the classical amplification field (with relative phase $\phi_{c}$),
respectively, while $\sigma_{rs}\equiv\left\vert r\right\rangle \left\langle
s\right\vert $ ($r$ and $s$ labeling the states $g$ or $e$). The vibrational
mode is decoupled from the remaining degrees of freedom of our model by
assuming the wave vectors $\overrightarrow{k}$ and $\overrightarrow{k}_{c}$ to
be perpendicular to $\overrightarrow{x}$; otherwise, a sufficiently small
Lamb-Dicke parameter keeps the motional state almost unchanged. Under this
assumption we arrive at the Hamiltonian%
\begin{equation}
H=\omega_{a}a^{\dagger}a+\frac{\omega_{0}}{2}\sigma_{z}+\left\{  ga\sigma
_{eg}+\Omega_{c}\operatorname{e}^{i\left(  \phi_{c}-\omega_{c}t\right)
}\sigma_{eg}+\mathrm{H{.c.}}\right\}  \text{,} \label{hamilto_partida}%
\end{equation}
which will be the starting point for our purposes. In the following, we show
how to engineer reservoirs suitable for obtaining nonadiabatic evolutions of
the internal ionic states.

\subsubsection{Nonadiabatic Evolution}

To engineer the appropriate interaction between the $TL$ ion and the cavity
mode we apply the unitary transformation given by $U_{1}=\exp\left[  -i\left(
\omega_{a}a^{\dagger}a+\omega_{c}\sigma_{z}/2\right)  t\right]  $, leading to
the Hamiltonian
\begin{equation}
H_{1}=\frac{\Delta_{c}}{2}\sigma_{z}+\left[  \left(  \left\vert \Omega
_{c}\right\vert \operatorname{e}^{i\phi_{c}}+g\operatorname{e}^{-i\delta_{a}%
t}a\right)  \sigma_{eg}+\mathrm{H{.c.}}\right]  ,\nonumber
\end{equation}
with $\Delta_{c}=\omega_{0}-\omega_{c}$ ($\delta_{a}=\omega_{a}-\omega_{c}$)
being the detuning between the atomic transition (cavity mode) and the laser
field. Moving to another frame of reference defined by the unitary
transformation $U_{2}=\exp\left\{  -i\left[  \left(  \Delta_{c}/2\right)
\sigma_{z}+\left\vert \Omega_{c}\right\vert \left(  \sigma_{eg}%
\operatorname{e}^{i\phi_{c}}+\sigma_{ge}\operatorname{e}^{-i\phi_{c}}\right)
\right]  t\right\}  ,$ the foregoing calculations are significantly
simplified. This procedure leads to the Hamiltonian%
\begin{align}
H_{2}  &  =g\operatorname{e}^{-i\delta_{a}t}a\left[  \lambda\left(
\sigma_{++}-\sigma_{--}\right)  +\right. \nonumber\\
&  \left.  \lambda_{+-}\operatorname{e}^{i2\xi t}\sigma_{+-}+\lambda
_{-+}\operatorname{e}^{-i2\xi t}\sigma_{-+}\right]  \operatorname{e}%
^{-i\phi_{c}}+\mathrm{H{.c.}}\text{,} \label{H2}%
\end{align}
where $\lambda=\sqrt{4-\chi^{2}}/4$, $\lambda_{\pm\mp}=\mp\left(  2\pm
\chi\right)  /4$, and $\xi=\sqrt{\left\vert \Omega_{c}\right\vert ^{2}%
+\Delta_{c}^{2}/4}$.\ The atomic operators are defined by $\sigma_{\pm\pm
}=\left\vert \pm\right\rangle \left\langle \pm\right\vert $, with $\left\vert
\pm\right\rangle =\frac{1}{2}\left(  \sqrt{2\pm\chi}\left\vert e\right\rangle
\pm\operatorname{e}^{-i\phi_{c}}\sqrt{2\mp\chi}\left\vert g\right\rangle
\right)  $ and $\chi=\Delta_{c}/\xi$. By assuming a large detuning between the
cavity mode and the laser field, such that $\delta_{a}\gg g$, together with
the additional choice $\delta_{a}=-2\xi$, we obtain under the RWA the
effective Hamiltonian%
\begin{equation}
H_{eff}=g_{eff}\left(  \operatorname{e}^{i\phi_{c}}\sigma_{+-}a^{\dagger
}+\operatorname{e}^{-i\phi_{c}}\sigma_{-+}a\right)  , \label{Hef2}%
\end{equation}
with $g_{eff}=g\left(  1-\chi/2\right)  /2$. The effective coupling thus
depends on the parameter $\chi$, whose value follows from the laser detuning
$\Delta_{c}$ which must be significantly smaller than the Rabi frequency
$\Omega_{c}$.

Next, observing that $\chi\in\left[  -2,2\right]  $, we focus on the values
$\chi\ll2$ which allow for an effective coupling of the same order of the
atom-cavity field coupling, i.e., $g_{eff}\left(  \chi\right)  \sim g$. Once
achieved the building of the interaction between the two-level system and the
cavity mode, we now take into account their interaction with a thermal
reservoir at temperature $T=0K$ through the master equation%
\begin{align}
\overset{\cdot}{\widetilde{\rho}^{(sa)}}  &  =-i\left[  H_{eff},\widetilde
{\rho}^{(sa)}\right] \nonumber\\
&  +\frac{\kappa}{2}\left(  2a\widetilde{\rho}^{(sa)}a^{\dagger}-a^{\dagger
}a\widetilde{\rho}^{(sa)}-\widetilde{\rho}^{(sa)}a^{\dagger}a\right)
\label{Master11}\\
&  +\frac{\gamma}{2}\left(  2\widetilde{\sigma}_{ge}\widetilde{\rho}%
^{(sa)}\widetilde{\sigma}_{eg}-\widetilde{\sigma}_{ee}\widetilde{\rho}%
^{(sa)}-\widetilde{\rho}^{(sa)}\widetilde{\sigma}_{ee}\right)  ,\nonumber
\end{align}
where again the supra index in $\widetilde{\rho}^{(sa)}$ denotes the density
operator for both system and field and the tilde is used to describe the
operators in the same representation of $H_{eff}$\textbf{,} i.e.,
$\widetilde{\mathcal{O}}=U_{2}^{\dagger}U_{1}^{\dagger}OU_{1}U_{2}$\textbf{.}
Here the constants\ $\kappa$ and $\gamma$ are the decay rates of the cavity
mode and the TL system, respectively. Now, except by the superoperator
describing the decay of the two-level system, we can see that Eq.
(\ref{Master11}) is equivalent to Eq. (\ref{master-geral}), with
$\lambda_{eff}=g_{eff}$ and $\mathfrak{O}=\operatorname{e}^{i\phi_{c}}%
\sigma_{+-}$ . As described above, assuming a strong decay rate $\kappa$ we
can adiabatically eliminate the cavity mode variables. This procedure is
needed to make clear what is the asymptotically protected atomic
superposition. Then, the dynamics of the reduced density operator for the $TL$
system results to be ($\widetilde{\rho}^{(s)}\equiv\widetilde{\rho}_{at}$)%
\begin{equation}
\overset{\cdot}{\widetilde{\rho}}_{at}\simeq\frac{\Gamma_{eng}}{2}\left(
2\sigma_{+-}\widetilde{\rho}_{at}\sigma_{-+}-\sigma_{--}\widetilde{\rho}%
_{at}-\widetilde{\rho}_{at}\sigma_{--}\right)  +\widetilde{\mathcal{L}%
}\widetilde{\rho}_{at}\text{,} \label{mestra_1laser}%
\end{equation}
so that $\Gamma_{eng}=4g_{eff}^{2}/\kappa$ represents the effective damping of
the engineered reservoir and $\widetilde{\mathcal{L}}\widetilde{\rho}_{at}%
$\textbf{ }is the last term in the r.h.s. of Eq. (\ref{Master11}) after
tracing over the cavity field variables. Note that the operator $\sigma_{+-}$
and consequently the protected state $\left\vert +\right\rangle $
($\sigma_{+-}\left\vert +\right\rangle =0$), exhibit no temporal dependences
in the convenient representation where we have described the evolution
(\ref{mestra_1laser}). However, as we shall see in the following, a coherent
nonadiabatic evolution is recovered in the interaction picture, where
$\left\vert +\left(  t\right)  \right\rangle =UU_{1}U_{2}\left\vert
+\right\rangle $ and $U=\exp\left[  i\left(  \omega_{a}a^{\dagger}a+\omega
_{0}\sigma_{z}/2\right)  t\right]  $, even under spontaneous decay, provided
that $\gamma\ll\Gamma_{eng}$. Here we note that the combined unitary
operations $UU_{1}U_{2}$ act on the Hilbert space of the atom and the mode.
However, once we are interested in the two-level system only and since the
field and atomic operators commute with each other, we are omitting, for
convenience, the corresponding state of the mode.

\textit{Ideal case:}\textbf{ }In the ideal case where $\gamma=0$, the solution
of Eq. (\ref{mestra_1laser}) leads to the $TL$ steady state $\widetilde{\rho
}_{at}=$ $\left\vert +\right\rangle \left\langle +\right\vert $. By returning
to the interaction picture we thus obtain $\rho_{at}=\left\vert +\left(
t\right)  \right\rangle \left\langle +\left(  t\right)  \right\vert $,
where\textbf{ }%
\begin{equation}
\left\vert +\left(  t\right)  \right\rangle =\frac{1}{2}\left(  \sqrt{2+\chi
}\left\vert e\right\rangle +\operatorname{e}^{-i\Phi(t)}\sqrt{2-\chi
}\left\vert g\right\rangle \right)  , \label{estado_prot}%
\end{equation}
with $\Phi(t)=\phi_{c}-\Delta_{c}t$. This state describes a nonadiabatic
evolution that depends on the detuning $\Delta_{c}$ between the atomic
transition and the classical field. Note that under the restriction $\chi\ll2$
imposed above, the states around the north pole of the Bloch sphere are not
allowed steady states.

\textit{Nonideal case: }To appreciate the robustness of the present reservoir
engineering technique, it is necessary to take into account the damping
stemming from the natural reservoir when $\gamma\neq0$. Here we will consider
the reservoir at zero temperature. The effect of finite temperature in the
fidelity of the protected state will be analyzed in the next Section. For
convenience, we will analyze such effects in the frame on which we have
defined $H_{1}$. To this end, we use $\widetilde{\rho}_{at}=U_{2}^{\dagger
}\rho_{at}U_{2}$ in Eq. (\ref{mestra_1laser}), which can thus be written as%
\begin{align}
\overset{\cdot}{\rho_{at}} &  =-i\left[  \xi\left(  \sigma_{++}-\sigma
_{--}\right)  ,\rho_{at}\right]  \nonumber\\
&  +\frac{\Gamma_{eng}}{2}\left(  2\sigma_{+-}\rho_{at}\sigma_{-+}-\sigma
_{--}\rho_{at}-\rho_{at}\sigma_{--}\right)  \label{mestra_1laser_U}\\
&  +\frac{\gamma}{2}\left(  2\sigma_{ge}\rho_{at}\sigma_{eg}-\sigma_{ee}%
\rho_{at}-\rho_{at}\sigma_{ee}\right)  \text{.}\nonumber
\end{align}
By projecting Eq. (\ref{mestra_1laser_U}) on the atomic basis we find the
following set of equations corresponding to the matrix elements $\rho
_{at}^{++}\equiv\left\langle +\right\vert \rho_{at}\left\vert +\right\rangle $
, $\rho_{at}^{+-}\equiv\left\langle +\right\vert \rho_{at}\left\vert
-\right\rangle $, and $\rho_{at}^{-+}\equiv\left\langle -\right\vert \rho
_{at}\left\vert +\right\rangle $:
\begin{align*}
\overset{\cdot}{\rho}_{at}^{++} &  =\left.  \Gamma_{eng}+\widetilde{\gamma
}\left(  \chi^{2}-2\right)  -\left[  \Gamma_{eng}+2\widetilde{\gamma}\left(
\chi^{2}+4\right)  \right]  \rho_{at}^{++}\right.  \\
&  \left.  -\widetilde{\gamma}\chi\sqrt{4-\chi^{2}}\left(  \rho_{at}^{+-}%
+\rho_{at}^{-+}\right)  \right.  \text{,}\\
\overset{\cdot}{\rho}_{at}^{+-} &  =\left.  \widetilde{\gamma}\left(  \chi
^{2}-2\right)  -2\widetilde{\gamma}\chi\sqrt{4-\chi^{2}}\rho_{at}^{++}\right.
\\
&  \left.  -\left[  \frac{\Gamma_{eng}}{2}+2\widetilde{\gamma}\left(  \chi
^{2}-12\right)  -i2\xi\right]  \rho_{at}^{+-}\right.  \\
&  \left.  +\widetilde{\gamma}\sqrt{4-\chi^{2}}\rho_{at}^{-+}\right.  \text{,}%
\end{align*}
with $\widetilde{\gamma}=\gamma/16$. Imposing the condition $\overset{\cdot
}{\rho}_{at}=0$, we can also determine the asymptotic solutions for $\rho
_{at}^{++}$ and $\rho_{at}^{+-}$, greater simplified considering a large
cooperative parameter $\mathcal{C}=g^{2}/\gamma\kappa\simeq\Gamma_{eng}%
/\gamma\gg1$, given by%
\begin{align*}
\rho_{++} &  =1-\epsilon_{++}\text{ ,}\\
\rho_{+-} &  =-i\epsilon_{+-}\text{,}%
\end{align*}
where $\epsilon_{++}=\left(  \gamma/\Gamma_{eng}\right)  \left[  \left(
2+\chi\right)  /8\right]  ^{2}$ and $\epsilon_{+-}=\gamma/4g_{eff}$. Under the
condition $\Gamma_{eng}/\gamma\gg1$, we see that both $\epsilon_{++}$ and
$\epsilon_{+-}$ are much smaller than unity. Actually, taking into account the
noise effects introduced by the reservoir, the steady state is approximately
described by $\rho_{at}\simeq\left\vert +\right\rangle \left\langle
+\right\vert $. The net effect of the noise introduced by the atomic decaying
mechanism is computed through the fidelity%
\[
\mathcal{F}=\mathrm{Tr}\left[  \left\vert +\right\rangle \left\langle
+\right\vert \rho_{at}\right]  =1-\epsilon_{++}\text{.}%
\]

The approximation leading to Eq. (\ref{mestra_1laser_U}) is better as higher
the decay rate $\kappa$; however, for the dynamics of the $TL$ system to be
driven by the engineered reservoir, the magnitude of $\kappa$ must be chosen
within the restriction $\mathcal{C}\gg1$. Within the optical regime
\cite{Hood}, for example, where $g\simeq7\times10^{8}$ s$^{-1}$ and
$\gamma\simeq2\times10^{7}$ s$^{-1}$, we obtain for a cavity decay constant
$\kappa\simeq3g$, the strength $\kappa/\gamma\simeq10$ and a fidelity around
$96\%$. Therefore, under the excellent approximation $\epsilon_{++}\ll1$ and
reversing the unitary transformation $R(t)$, the state $\left\vert
+\right\rangle $, written in the interaction picture Eq. (\ref{estado_prot})
allows for a nonadiabatic\ coherent evolution of a $TL$ system under
spontaneous decay, which can be manipulated through the parameter $\Delta_{c}%
$. Such an evolution corresponds to flips in the atomic states, representing
trajectories on different planes parallel to the equator on the Bloch sphere,
governed by the Hamiltonian $H=\Omega_{c}\sigma_{eg}\operatorname*{e}%
\nolimits^{-i(\Delta_{c}t-\phi_{c})}+H.c.$.

\section{Numerical results and temperature effects}

To validate the approximations carried out in the previous Section, we proceed
to numerically evaluate the full Hamiltonian Eq. (\ref{hamilto_partida}), in
the interaction picture, taking into account the thermal reservoir for both
the atom and the field, such that%
\begin{align}
\overset{\cdot}{\rho^{(sa)}}  &  =-i\left[  V,\rho^{(sa)}\right] \nonumber\\
&  +\frac{\kappa\left(  \overline{n}_{a}+1\right)  }{2}\left(  2a\rho
^{(sa)}a^{\dagger}-a^{\dagger}a\rho^{(sa)}-\rho^{(sa)}a^{\dagger}a\right)
\nonumber\\
&  +\frac{\kappa\overline{n}_{a}}{2}\left(  2a^{\dagger}\rho^{(sa)}%
a-aa^{\dagger}\rho^{(sa)}-\rho^{(sa)}aa^{\dagger}\right) \nonumber\\
&  +\frac{\gamma\left(  \overline{n}_{s}+1\right)  }{2}\left(  2\sigma
_{ge}\rho^{(sa)}\sigma_{eg}-\sigma_{ee}\rho^{(sa)}-\rho^{(sa)}\sigma
_{ee}\right) \nonumber\\
&  +\frac{\gamma\overline{n}_{s}}{2}\left(  2\sigma_{eg}\rho^{(sa)}\sigma
_{ge}-\sigma_{gg}\rho^{(sa)}-\rho^{(sa)}\sigma_{gg}\right)  \text{,}
\label{MasterT}%
\end{align}
where%
\[
V=ga\sigma_{eg}\operatorname{e}^{-i\delta_{a}t}+\Omega_{c}\sigma
_{eg}\operatorname{e}^{i\left(  \phi_{c}-\Delta_{c}t\right)  }+\mathrm{H{.c.}%
}\text{,}%
\]
and $\overline{n}_{a}$ and $\overline{n}_{s}$ are the average number of
thermal photons for the bosonic field and atomic reservoirs, respectively.

\begin{figure}[th]
\centering
\includegraphics[width=1\linewidth]{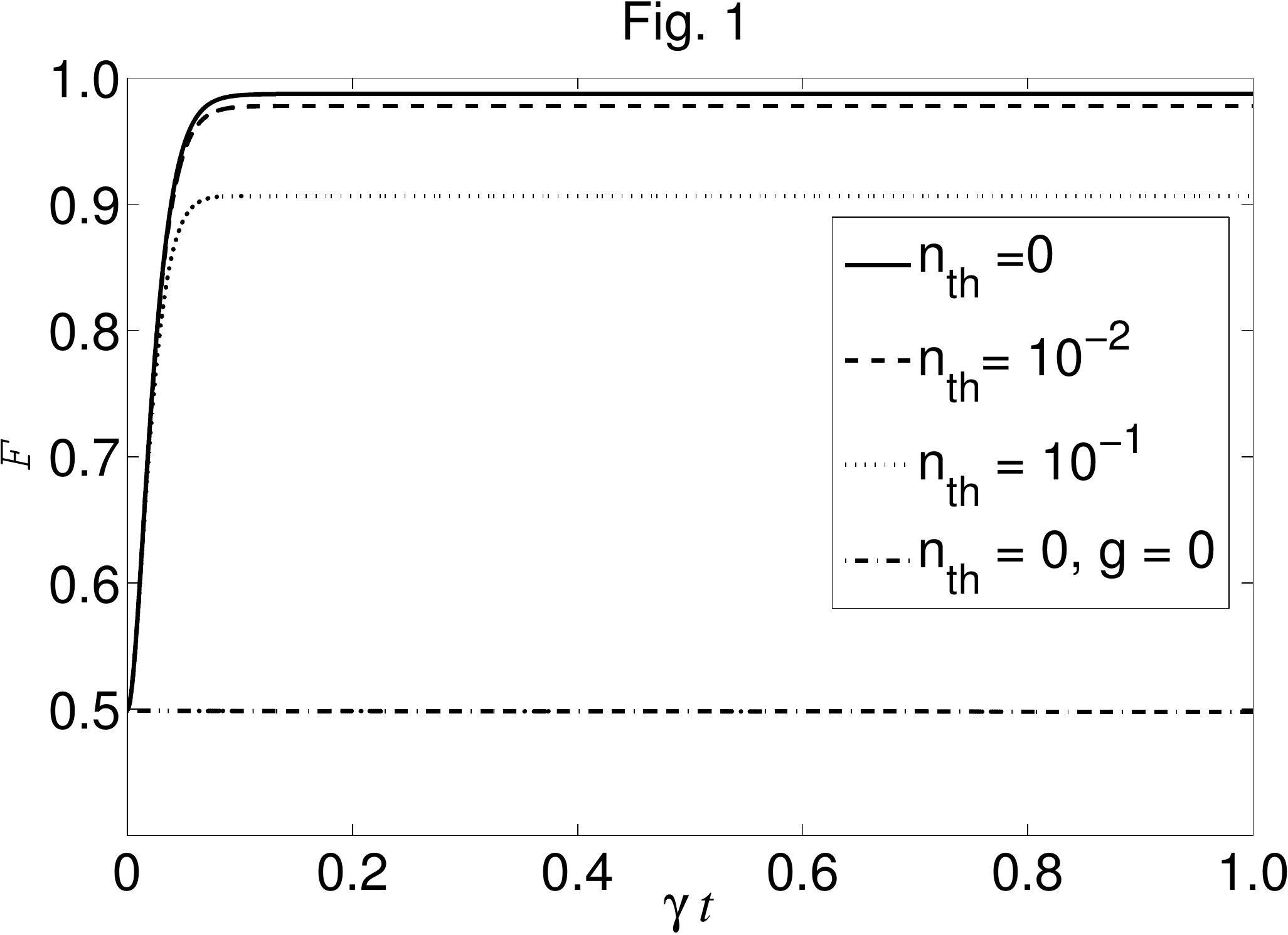}\caption{Evolution of the
numerically computed fidelity for both cases of absolute zero (solid line) and
finite temperature, with the average number of thermal photons $\overline
{n}=0.01$ (dashed line) and $\overline{n}=0.1$ (dotted line). The constant
value $\mathcal{F}=0.5$ indicated by the dashed-dotted line refers to the case
where the coupling $g$ between the atomic system and the cavity mode is null.}%
\label{Fig1}%
\end{figure}

We emphasize that, differently from the master equation (\ref{Master11}) which
has been derived under the RWA approximation leading to the effective
Hamiltonian (\ref{Hef2}), the above master equation (\ref{MasterT}) describes
exactly the dynamics of the whole system thus allowing us to quantify the
errors introduced by our approximations -- RWA in Eq. (\ref{Hef2}) and
adiabatic elimination of the field variables in Eq. (\ref{mestra_1laser}).
Since we are assuming the atomic frequency near the resonance with the cavity
field frequency, both their reservoirs will have the same average thermal
photons, such that from now on we take $\overline{n}_{a}$ $=$ $\overline
{n}_{s}$. By projecting the above equation in the Fock $\left\{  \left\vert
n\right\rangle \right\}  $ and electronic $\left\{  \left\vert e\right\rangle
,\left\vert g\right\rangle \right\}  $ bases, we are lead to an infinity set
of coupled equations for the matrix elements. To solve numerically this system
of infinity coupled differential equations we must truncate the Fock basis
somewhere. The strong decay rate $\kappa$ allows us to safely do it since the
matrix elements corresponding to highly excited Fock states are virtually
zero. We then numerically solve the master equation (\ref{MasterT}) following
the method presented in \cite{master-matlab}. Since we are interested in the
evolution of the atomic two-level system only, we trace numerically over the
bosonic field variables. Therefore, after solving the full master equation we
end up with a density matrix for the two-level system with elements
\begin{equation}
\rho_{at}^{ij}\left(  t\right)  =\sum_{n=0}^{N}\rho_{n,n}^{ij}\left(
t\right)  \text{,} \label{Red}%
\end{equation}
where $\rho_{at}^{ij}\equiv\left\langle i\right\vert \rho_{at}\left\vert
j\right\rangle $.

Here we analyze the robustness of the protected atomic superposition given by
Eq. (\ref{estado_prot}). Our strategy consists in computing the robustness of
the protected state under thermal effects considering the following different
regimes: i) $\Phi(t)=$\textit{ constant}, which corresponds to the static
case, ii) by allowing $\Phi(t)$ to slowly vary with time, corresponding to
$\Delta_{c}/\Gamma_{eng}\ll1$, and iii) allowing $\Delta_{c}(t)$ to rapidly
vary in time, corresponding to $\Delta_{c}/\Gamma_{eng}>1$. The robustness of
the protected state is computed through the fidelity $\mathcal{F}%
(t)=\mathrm{Tr}\left[  \left\vert +(t)\right\rangle \left\langle
+(t)\right\vert \rho_{at}(t)\right]  $ where $\rho_{at}(t)$ follows from the
numerical solution of the full master equation (\ref{MasterT}). In units of
$\gamma$, we assumed the reasonable decay rates $\kappa=10^{2}\gamma$,
$g=10^{2}\gamma$ and $\Omega_{c}=2\times10^{3}\gamma$.

Starting with the static case, we consider $\Delta_{c}(t)=0$ with constant
$\phi_{c}$, such that $\left\vert +\right\rangle =\frac{1}{\sqrt{2}}\left(
\left\vert e\right\rangle +\operatorname{e}^{i\phi_{c}}\left\vert
g\right\rangle \right)  $. Assuming $\phi_{c}=0$ and the $TL$ atom prepared in
the ground state, in Fig. 1 we present the numerical results of the fidelity
against the parameter $\gamma t$ for both cases of absolute zero (solid line)
and finite temperature, with the average number of thermal photons
$\overline{n}=0.01$ (dashed line) and $\overline{n}=0.1$ (dotted line). As
expected, we verify that the temperature effects reduce substantially the
fidelity of the atomic superposition. To stress the effectiveness of our
protocol, we call the attention to the case where the coupling $g$ between the
atomic system and the cavity mode is null. In this case we observe that the
fidelity remains $\mathcal{F}=0.5$ (dashed-dotted line) showing that the
cavity mode is a crucial ingredient, together with the classical pumping, to
protect the desired superposition state.

\begin{figure}[th]
\centering
\includegraphics[width=1\linewidth]{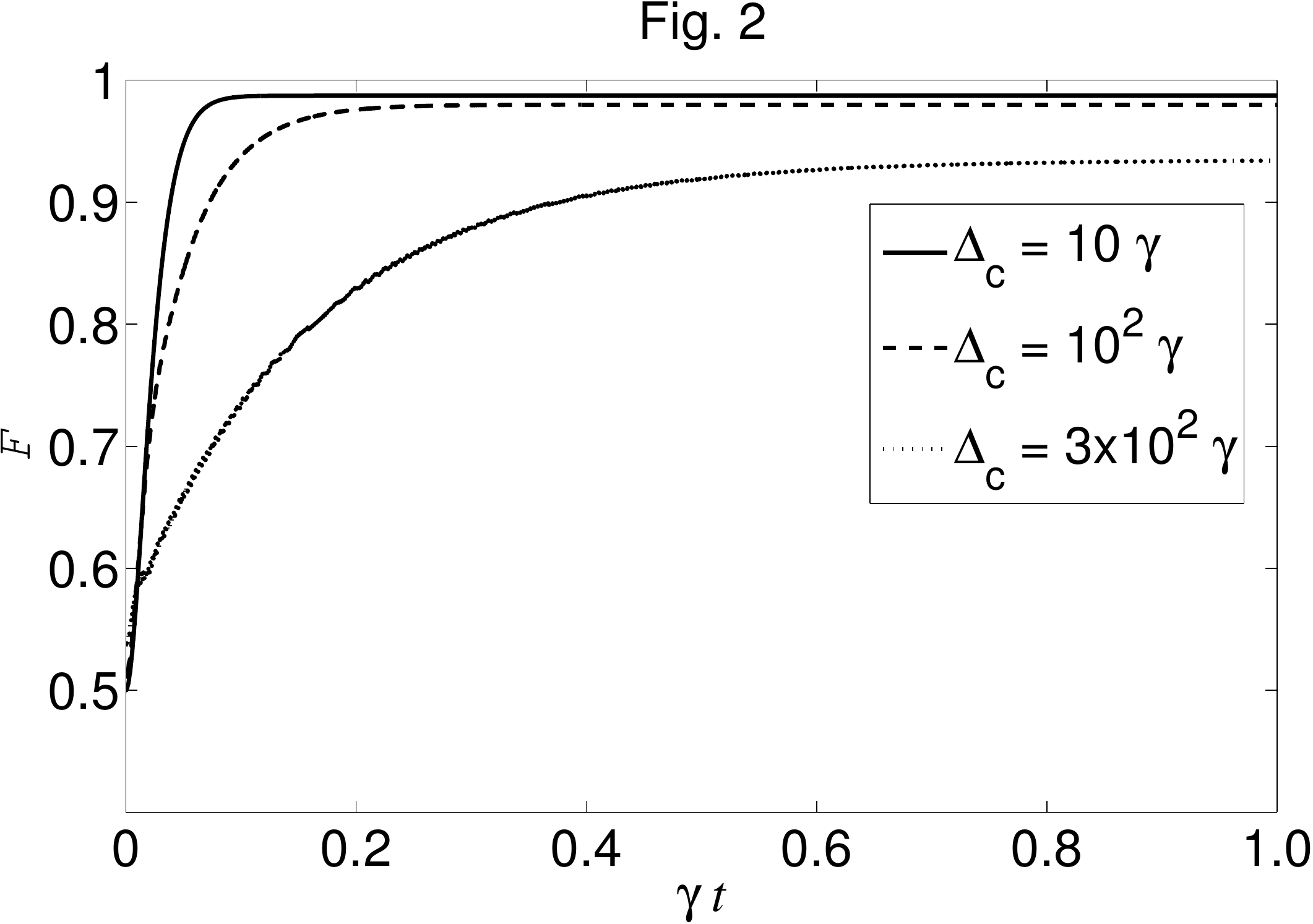}\caption{Fidelity of the
protected state $\left\vert +\left(  t\right)  \right\rangle $ undergoing
adiabatic and nonadiabatic evolutions at zero temperature. We consider the
values $\Delta_{c}/\Gamma_{eng}\simeq0.1$ (solid line), $\Delta_{c}%
/\Gamma_{eng}\simeq1$ (dashed line), and\ finally, $\Delta_{c}/\Gamma
_{eng}\simeq3$\ (dotted line).}%
\label{Fig2}%
\end{figure}

In Fig. 2 we plot the fidelity of the protected state $\left\vert +\left(
t\right)  \right\rangle $ undergoing slow and fast evolutions. We have
considered three distinct values for the detuning between the atom and the
classical field: $\Delta_{c}=10\gamma$, leading to a slow evolution with
$\Delta_{c}/\Gamma_{eng}\simeq0.1$ (solid line), $\Delta_{c}=10^{2}\gamma$,
departing from the adiabatic regime with $\Delta_{c}/\Gamma_{eng}\simeq1$
(dashed line), and\ finally $\Delta_{c}=3\times10^{2}\gamma$, with $\Delta
_{c}/\Gamma_{eng}\simeq3$\ (dotted line). All these curves in Fig. 2 were
plotted considering $\overline{n}=0$.\ As we are concerned with slow and fast
evolutions dictated by the time varying parameter $\Phi(t)=\phi_{c}-\Delta
_{c}t$, we focused our attention in $\Delta_{c}(t)$ taking $\phi_{c}=0$. We
stress that when the condition $\Delta_{c}/\Gamma_{eng}\ll1$ is weakened,
meaning that some parameters are rapidly varying in time, we found that
although the equilibrium is reached more slowly, the fidelity does not drop
off quickly, attaining a value about $0.9$ even for $\Delta_{c}/\Gamma
_{eng}\simeq3$, corroborating again the effectiveness of our protocol.

\begin{figure}[th]
\centering
\includegraphics[width=1\linewidth]{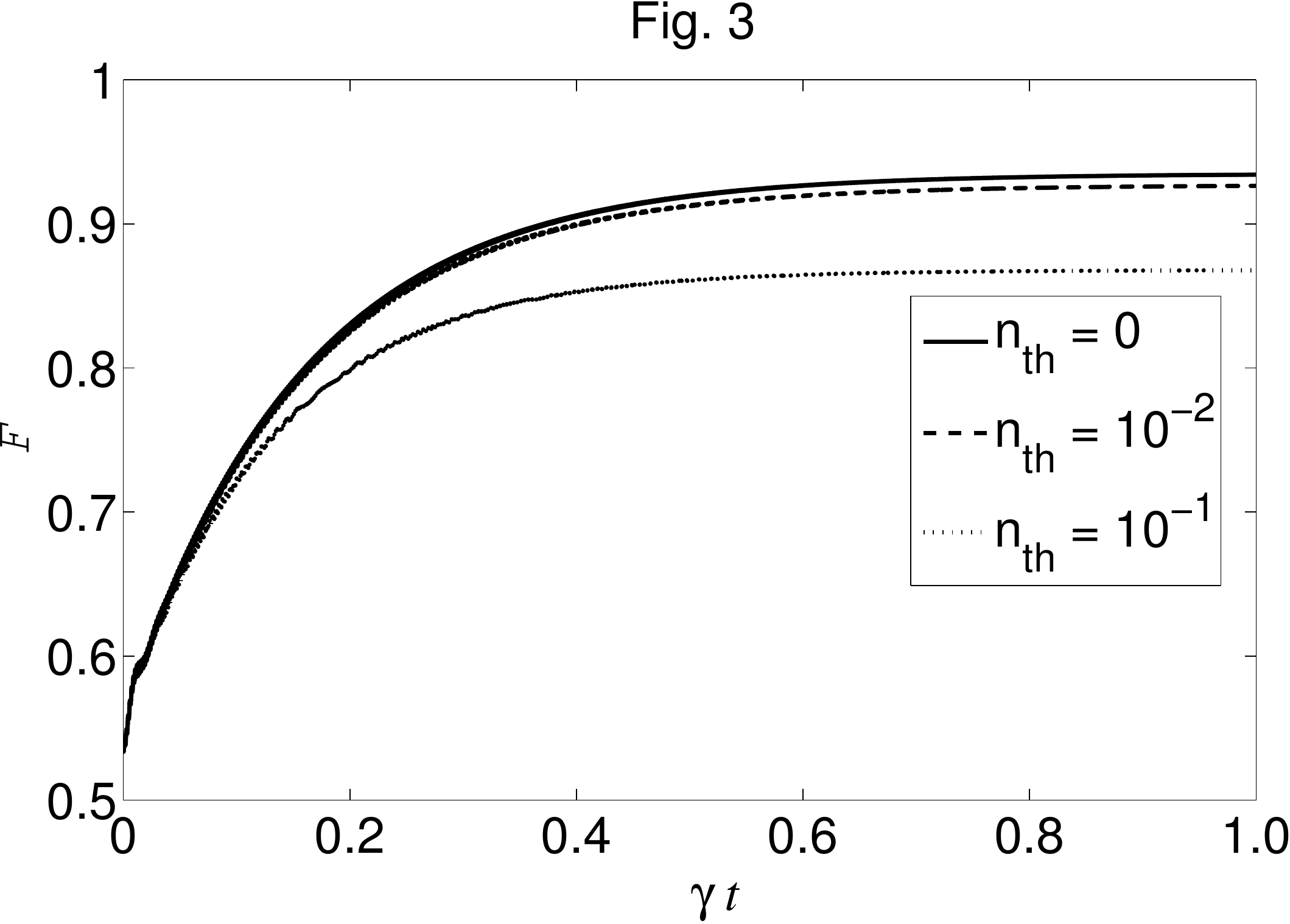}\caption{Fidelity of the
protected state $\left\vert +\left(  t\right)  \right\rangle $ undergoing
nonadiabatic evolutions, with $\Delta_{c}=3\times10^{2}\gamma$ ($\Delta
_{c}/\Gamma_{eng}\simeq3$), for absolute zero (solid line) and finite
temperature, with the average number of thermal photons $\overline{n}=0.01$
(dashed line) and $\overline{n}=0.1$ (dotted line).}%
\label{Fig3}%
\end{figure}

In order to see the temperature effects in the nonadiabatic evolutions, in
Fig. 3 we draw the curves of the fidelity for $\Delta_{c}=3\times10^{2}\gamma$
($\Delta_{c}/\Gamma_{eng}\simeq3$) for different values of mean number of
thermal photons, i.e., for absolute zero (solid line), $\overline{n}=0.01$
(dashed line) and $\overline{n}=0.1$ (dotted line). Again, as in the static
case (Fig. 1), we observe that the fidelity decreases with the increase of the
mean number of thermal photons, showing that the temperature is the most
important source of decoherence in our present protocol. This conclusion is
also drawn from Fig.4, where a functional dependence of the fidelity of the
protected state $\left\vert +\left(  t\right)  \right\rangle $ undergoing
nonadiabatic evolutions against the reservoirs mean photon number is
displayed. From this figure, where we have used $g=10\gamma$ (solid line),
$20\gamma$(dashed line), $50\gamma$ (dotted) and $10^{2}\gamma$ (dashed-dotted
line), we note that the fidelity decays faster with the increasing number of
thermal photons, as we have mentioned above.

\begin{figure}[th]
\centering
\includegraphics[width=1\linewidth]{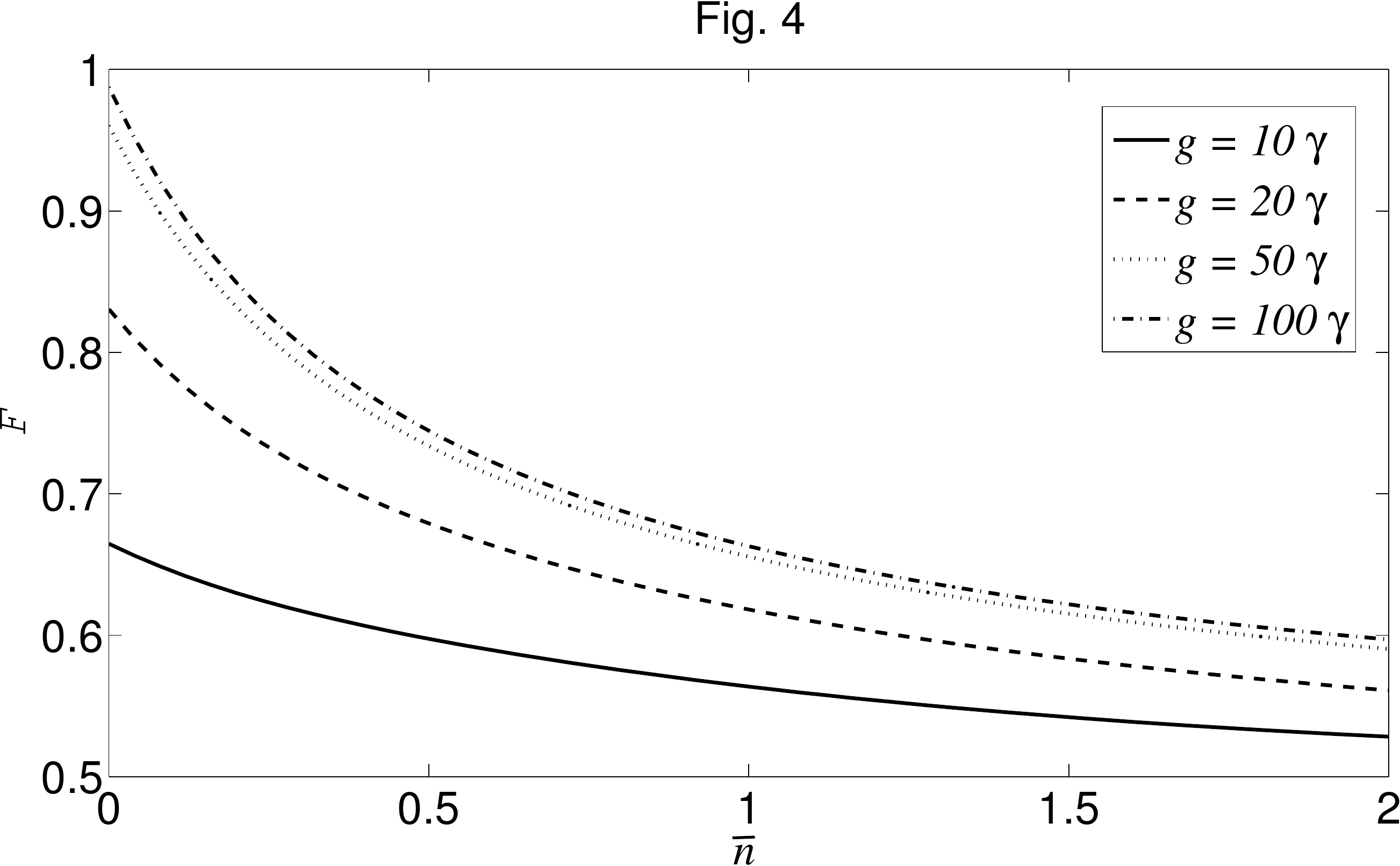}\caption{Fidelity of the
protected state $\left\vert +\left(  t\right)  \right\rangle $ undergoing
nonadiabatic evolutions, versus the average thermal photons $\overline{n}$
characterizing the finite temperatures with $g=10\gamma$ (solid line),
$20\gamma$(dashed line), $50\gamma$ (dotted) and $10^{2}\gamma$ (dashed-dotted
line).}%
\label{Fig4}%
\end{figure}

\section{Concluding Remarks}

In this paper we deepened the analysis of the engineering reservoir technique
we have proposed in Ref. \cite{PRLFabiano09}. This proposal has been
accomplished by deriving a time-dependent master equation which leads to a
decoherence-free evolving superposition state which can be nonadiabatically
controlled by the system-reservoir parameters. In the present contribution we
have provided a general recipe to engineer arbitrary effective reservoirs for
a fermionic system by manipulating its interaction with a bosonic mode.

More specifically, we showed how to protect a superposition state of a
two-level ion trapped into a leaky cavity. The robustness of our scheme was
analyzed by means of the fidelity considering the case where the protected
state does not depend on time, as well as the case of slowly and rapidly time
varying evolutions. To support the approximations used to derive our
analytical results, we have numerically solved the full master equation,
obtaining an excellent agreement with the approximations we have carried out.
In the present contribution we also analyzed the temperature effects of the
reservoirs on the fidelity of the static and time varying protected states. We
concluded that the temperature is the most important source of decoherence in
our present protocol.

We hope that this contribution can be useful for information processing with
trapped ions inside optical cavities, for example for implementing Deutsch
algorithm \cite{deutsch, eduardo} or universal dissipative quantum computing
\cite{cirac-naturephysics}.

We wish to express our thanks to FAPESP, FAPEMIG, CAPES, CNPq, and the
Brazilian National Institute of Science and Technology for Quantum Information
(INCT-IQ), for the financial support.

\end{document}